\documentclass[%
 reprint,
superscriptaddress,
showpacs,
 amsmath,amssymb,
 aps,prl
]{revtex4-1}
\usepackage{epsfig}
\usepackage{epstopdf}
\usepackage{graphicx}
\usepackage{dcolumn}
\usepackage{bm}
\usepackage{amsmath}
\usepackage{array}
\usepackage{color}

\begin{document}

\setlength{\belowdisplayskip}{0pt} \setlength{\belowdisplayshortskip}{0pt}
\setlength{\abovedisplayskip}{0pt} \setlength{\abovedisplayshortskip}{0pt}
\title{Photon-efficient quantum cryptography with pulse-position modulation }

\author{Tian Zhong}
\email{tzhong@caltech.edu}
\affiliation{Research Laboratory of Electronics, Massachusetts Institute of Technology, 77 Massachusetts Avenue, Cambridge, Massachusetts 02139, USA}

\author{Feihu Xu}
\affiliation{Research Laboratory of Electronics, Massachusetts Institute of Technology, 77 Massachusetts Avenue, Cambridge, Massachusetts 02139, USA}

\author{Zheshen Zhang}
\affiliation{Research Laboratory of Electronics, Massachusetts Institute of Technology, 77 Massachusetts Avenue, Cambridge, Massachusetts 02139, USA}
\author{Hongchao Zhou}
\affiliation{Research Laboratory of Electronics, Massachusetts Institute of Technology, 77 Massachusetts Avenue, Cambridge, Massachusetts 02139, USA}

\author{Alessandro Restelli}
\affiliation{Joint Quantum Institute, University of Maryland and National Institute of Standards and Technology, Gaithersburg, Maryland 20899, USA }
\author{Joshua C. Bienfang}
\affiliation{Joint Quantum Institute, University of Maryland and National Institute of Standards and Technology, Gaithersburg, Maryland 20899, USA }

\author{Ligong Wang}
\affiliation{Research Laboratory of Electronics, Massachusetts Institute of Technology, 77 Massachusetts Avenue, Cambridge, Massachusetts 02139, USA}

\author{Gregory W. Wornell}
\affiliation{Research Laboratory of Electronics, Massachusetts Institute of Technology, 77 Massachusetts Avenue, Cambridge, Massachusetts 02139, USA}
\author{Jeffrey H. Shapiro}
\affiliation{Research Laboratory of Electronics, Massachusetts Institute of Technology, 77 Massachusetts Avenue, Cambridge, Massachusetts 02139, USA}
\author{Franco N. C. Wong}
\email{ncw@mit.edu}
\affiliation{Research Laboratory of Electronics, Massachusetts Institute of Technology, 77 Massachusetts Avenue, Cambridge, Massachusetts 02139, USA}

\date{\today}

\begin{abstract}
The binary (one-bit-per-photon) encoding that most existing quantum key distribution (QKD) protocols employ puts a fundamental limit on their achievable key rates, especially under high channel loss conditions associated with long-distance fiber-optic or satellite-to-ground links. Inspired by the pulse-position-modulation (PPM) approach to photon-starved classical communications, we design and demonstrate the first PPM-QKD, whose security against collective attacks is established through continuous-variable entanglement measurements that also enable a novel decoy-state protocol performed conveniently in post processing. We achieve a throughput of 8.0 Mbit/s (2.5 Mbit/s for loss equivalent to 25 km of fiber) and secret-key capacity up to 4.0 bits per detected photon, thus demonstrating the significant enhancement afforded by high-dimensional encoding. These results point to a new avenue for realizing high-throughput satellite-based or long-haul fiber-optic quantum communications beyond their photon-reception-rate limits.
\end{abstract}

\maketitle

\noindent Quantum key distribution (QKD) \cite{bb84,E91} allows two remote users, Alice and Bob, to  create a shared random key while precluding an eavesdropper (Eve) from obtaining any meaningful amount of information. To date, numerous QKD protocols have demonstrated robust extraction of secret keys over 200 km of fiber, and satellite-to-ground links, with secret-key rates up to the order of 10 bit/s \cite{BB8410,shields,COW, COW15, DPS, DPS07, entQKD, entQKD2,freespace}. These protocols, namely decoy-BB84 \cite{BB8410,shields}, coherent one-way (COW) \cite{COW, COW15}, differential phase-shift (DPS) \cite{DPS,DPS07}, and entanglement-based QKD \cite{entQKD,entQKD2}, share a common binary (qubit) encoding, with photon information efficiency (PIE) of at most 1 bit per detected photon. Their secret-key rates are severely limited by the photon flux at the receiver, typically on the order of Hz for long-distance transmission~\cite{freespace}. QKD usually operates under photon-starved conditions in which the receiver's photon detection rate is dramatically lower than the transmitter's photon generation rate owing to propagation loss and less-than-unity efficiencies in single-photon detectors \cite{SPD}.

For classical optical communications in high loss scenarios, a solution to photon-starved reception is to utilize pulse-position modulation (PPM)~\cite{Robinson}, a form of signal modulation in which $k$ message bits are encoded by transmitting a single pulse in one of $N=2^k$ possible time bins. PPM is particularly useful in free-space satellite communication links including the recent lunar laser-communication demonstration~\cite{lunar}, achieving high data rates with an excellent average-power efficiency. A quantum version of PPM could enable photon-efficient satellite-based or long-distance QKD with multiple bits encoded in a single photon, leading to secret-key rates much beyond the photon reception limit. Nevertheless, its demonstration remains elusive as it requires pulse-position modulation/demodulation performed at the single-photon level. Moreover, the nature of high-dimensional encoding \cite{Howell, Zhong, Zhang, Lee, Xie} of PPM-QKD precludes the use of conventional security proofs for qubit protocols \cite{Lo:1999,Shor:2000,Tomamichel,decoy}. A new PPM-compatible protocol with tailored error-correction and finite-key analysis is needed to reap its potential advantages.

Here we demonstrate a hybrid QKD architecture that uses weak amplified spontaneous emission (ASE) pulses for PPM prepare-and-measure key generation, and a continuous-variable entanglement probe to establish security via near-unity-visibility Franson interferometry~\cite{Franson, ndc}. We combine the analysis of decoy-state protocols~\cite{decoy} in discrete-variable QKD, which estimates the fraction of single-photon detections, with the security proof against collective attacks in continuous-variable QKD \cite{CVQKD,patron} to bound Eve's Holevo information. Our proof-of-principle PPM-QKD experiment realized a high PIE up to 4.0 secret bits per detected photon and achieved a maximum throughput of 2.5 (8.0) Mbit/s at a propagation loss equivalent to that of 25 ($\approx$ 0) km fiber distance. This high rate was obtained with considerably fewer detected photons than what would be required for a comparable rate with the state-of-the-art decoy-state BB84 protocol. Similar to PPM usage in classical communications, we also show enhancements of PPM-QKD throughput beyond the photon reception limit at increasing channel losses, thus demonstrating a useful technique for high-rate satellite-based QKD or long-haul fiber-optic quantum networks.

\begin{figure}[ht]
\includegraphics[width=0.45\textwidth]{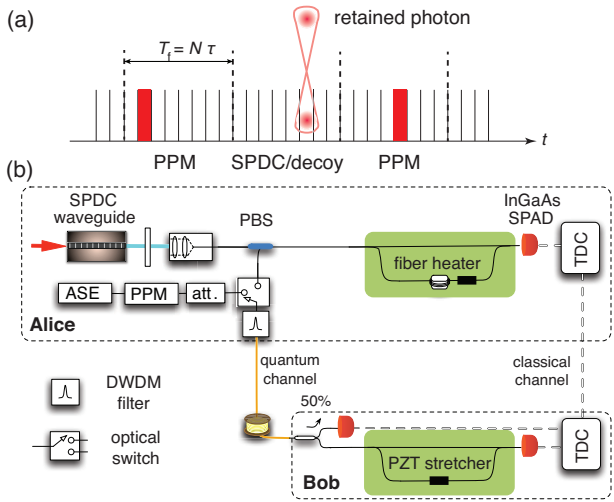}
\caption{(color online). Schematic of PPM-QKD: (a) time segment of several time frames $T_f$ with ASE and SPDC signal photons; (b) experimental setup with a 50:50 fiber beam splitter used by Bob for key generation and Franson measurements. See text for additional details. PBS: polarizing beam splitter, TDC: time-to-digital converter, SPAD: single-photon avalanche diode, att: attenuator.}\label{f1}
\end{figure}

The basic PPM-QKD scheme is illustrated in Fig.~\ref{f1}(a), with the experimental setup shown in Fig.~\ref{f1}(b). PPM key bits are encoded by modulating weak ASE light in time frames of duration $T_f$ each comprising $N$ $\tau$-duration time bins. We randomly insert additional $T_f$-duration frames, containing time-energy entangled photons generated from continuous-wave (cw) spontaneous parametric downconversion (SPDC), into the stream of PPM frames. These SPDC frames serve two critical functions. First, they allow the recent security proof for high-dimensional temporal encoding via Franson interferometry~\cite{Zhang} to be applied to PPM-QKD. Second, they provide a decoy state---by using different mean photon numbers for the ASE and SPDC light---from which the single-photon component in the ASE PPM can be determined.  Filtering is used to match the spectra of the ASE and SPDC light so that they present indistinguishable thermal-state statistics to Eve.

For each time frame, Alice randomly chooses between sending ASE or SPDC light to Bob through an optical fiber that is subject to Eve's attack. Alice and Bob share a publicly-synchronized clock to align their time bins and $N$-bin time frames. Bob randomly measures the photon arrival time either directly, for extracting symbols of $k=\log_2N$ bits, or after his arm of the Franson interferometer. Whenever Alice transmits SPDC signal light she measures her retained idler beam through her arm of the Franson interferometer to establish security. After many uses of the quantum channel, Alice publicly announces her choice of the sources used for each frame. For ASE frames, Bob first sifts the raw symbols from those frames, then proceeds to error correction and privacy amplification.

In Alice's random selection of PPM (for key generation) or SPDC (for Franson interference) light to send to Bob, we employ an asymmetric selection ratio, favoring PPM, in order to achieve a higher key generation rate than that obtainable with a 50:50 ratio. Eve's inability to distinguish which source Alice has used allows her accessible information to be bounded by visibility measurement from a single Franson interferometer. The PIE (bits per photon) is given by~\cite{CVQKD,patron}
\begin{equation}  \label{eq:deltaI}
\Delta I_{\rm AB}^{N}=\beta I_{\rm AB}^{N}-\chi^{\rm E},
\end{equation}
where $\beta$ is the reconciliation efficiency, $ I_{\rm AB}^{N} $ is Alice and Bob's Shannon information (SI) for a frame size $N$, and $\chi^{\rm E}$ is Eve's Holevo information for a collective Gaussian attack considering finite key lengths \cite{CVQKD}. Our finite-key analysis follows~\cite{HDFK} and its details are given in the Supplemental Material \cite{supplemental}. To bound Eve's Holevo information, Alice and Bob monitor the Franson interference visibility $V$, which is linked to the two-photon frequency anti-correlation by $ V=\langle\cos [(\hat{\omega}_{\rm A}-\hat{\omega}_{\rm B})\Delta T]\rangle $, where $\Delta T$ is the propagation delay between the interferometer's long and short paths, and $\hat{\omega}_{\rm A}(\hat{\omega}_{\rm B})$ is the frequency operator measuring the zero-mean detuning of Alice's (Bob's) photon at the frequency $\omega_{\rm p}/2+\omega_{\rm A}$ $(\omega_{\rm p}/2-\omega_{\rm B})$. Here $\omega_{\rm p}$ is the center frequency of the SPDC pump laser. Measuring the degradation of the visibility from its theoretical value $V^{\rm th}$ yields an upper bound on $\chi^{\rm E}$ (see Supplemental Material \cite{supplemental} or \cite{Zhong} for details).

The overall secret-key rate (bit/s) is given by~\cite{Darius}
\begin{equation}  \label{eq:key}
R=R_{f}[\underline{Q_{1}}(n_{\text{R}}-\overline{\chi^{\rm E}_{1}})-Q_{\mu}(n_{\text{R}}-\beta I_{\rm AB}^N)],
\end{equation}
where $R_{f}$ is the frame rate per second, $n_{\text{R}}$ is the number of random bits shared between Alice and Bob after error correction, $Q_{\mu}$ is the overall gain (the probability for Bob to obtain a detection event in a frame), $\underline{Q_{1}}$ and $\overline{\chi^{\rm E}_{1}}$ are, respectively, the lower bound on the single-photon gain and the upper bound of Eve's Holevo information for frames in which Alice sends a single-photon state and Bob obtains a detection. $Q_{\mu}$ can be measured directly from experiment, while $\underline{Q_{1}}$ and $\overline{\chi^{\rm E}_{1}}$ can be estimated from our novel decoy-state method that can be performed in post processing (See Supplemental Material \cite{supplemental} for details). 


\begin{figure*}[tb]
\includegraphics[width=.95\textwidth]{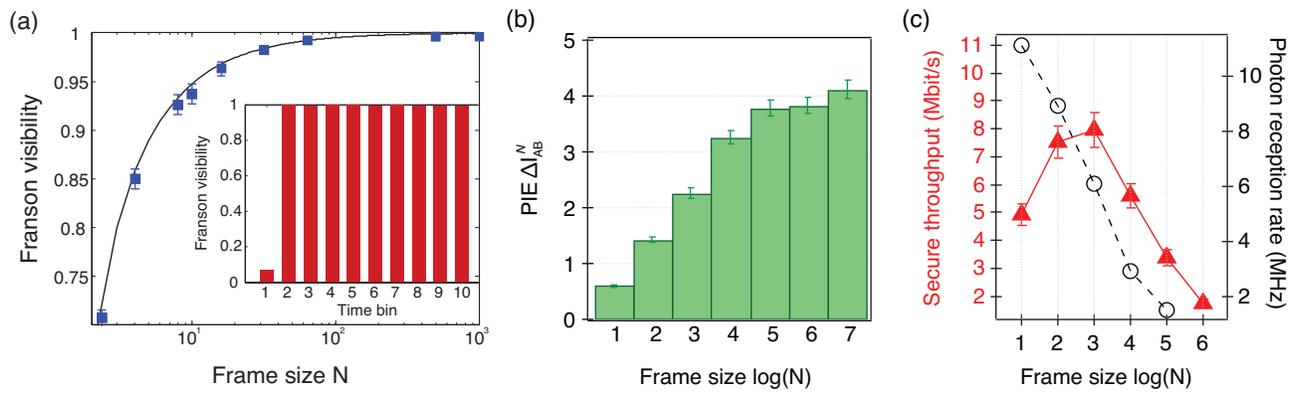}
\caption{(color online). (a) Measured Franson visibilities as a function of frame size $N$. Inset shows time-bin resolved Franson visibilities for $N=10$. Near-unity visibilities are observed in all bins except the first one. (b) Optimized PIE with different frame sizes. Green bars are PIE after error correction and privacy amplification. Error bars are due to uncertainties in the Franson visibilities. (c) Short-distance PPM-QKD throughput and photon reception rate as functions of frame size.}\label{f2}
\end{figure*}

In our proof-of-principle experiment shown schematically in Fig.~\ref{f1}(b), the ASE light from an erbium-doped fiber amplifier was intensity modulated for PPM with random symbols from a 10 Gbit/s pattern generator. The duration of each PPM pulse was $\approx$100 ps, matching the detector gate width (see below). The ASE light was then attenuated to an average of $\mu_{\text{PPM}}$ photons in the occupied bin. For security, a type-II phase-matched single-spatial-mode periodically poled KTiOPO$_4$ waveguide generated a stream of high quality time-energy entangled photon pairs at 1560 nm via SPDC \cite{zhong12}, and was operated with a mean photon pair $\mu_{\rm SPDC}$ per bin. The orthogonally polarized signal and idler pairs were separated by a polarizing beam splitter with the idler photons directed to Alice's arm of the Franson interferometer.

Alice used an optical switch to route the ASE or SPDC signal light through a dense wavelength-division multiplexer (DWDM) filter to Bob. The DWDM filter had a Gaussian spectral shape with a 200 GHz (1.6 nm) bandwidth full-width at half-maximum, which matches that of the SPDC spectrum \cite{zhong12}. The ASE and SPDC signal light were thus spectrally identical after DWDM filtering. The optical switch was driven at a frame rate of 1.26/$N$ GHz and the time bin duration $\tau$ was 794 ps. We set a 7:3 ratio between PPM frames (with ASE light sent) and Franson frames (with SPDC signal light sent).  A higher ratio would favor higher QKD throughput, but also require longer integration times for Franson measurements. The protocol requires one or more true quantum random-number generators (QRNGs). Although only pseudo-random numbers were used, we note that QRNGs operating at Gbit/s have been developed \cite{RNG}.

Bob used a 50:50 beam splitter to passively choose between key generation and Franson security check.  To achieve near-unity Franson interferometric visibilities, nonlocal dispersion cancellation \cite{ndc} was implemented by applying a negative differential dispersion (using low dispersion LEAF fiber) in Bob's arm of the interferometer. The long--short fiber length difference of the Franson interferometer in Fig.~\ref{f1}(b) was fixed at $\tau$ = 794 ps.

The InGaAs single-photon avalanche diodes (SPADs) were operated in the self-differencing mode at a gating frequency of 1.26 GHz (794 ps time bin), with detection efficiencies of 18\% at $1560$ nm and an effective detection gate width of $\approx$100 ps \cite{Yuan, zhong12}. The SPADs were operated at 5$^{\circ}$\,C to minimize afterpulsing \cite{Restelli}. The dark count rates were $\approx8\times10^3$ counts per second at this temperature. Detection events were time stamped with time-to-digital converters (TDCs) whose time base was phase locked to the SPAD gating signals.

Alice and Bob's raw timing data were collected from the TDCs after every 50 s QKD session. For each PPM frame,  Bob's data were first parsed into $\log_2 N$ bit symbols. We then performed error correction using a custom code developed by Zhou \emph{et al.}~\cite{HCZ} for high-dimensional QKD (HDQKD) with blocks of 4000 symbols each, and outputted the error-corrected symbols. After calculating the PIE, corrected symbols were fed into the privacy amplification algorithm~\cite{PA} to obtain the final keys. The key length for each session was $5\times10^7$ symbols. The finite key length penalty $\Delta_{\rm FK}$ due to error correction ($\epsilon_{\rm EC}$), privacy amplification ($\epsilon_{\rm PA}$) and smooth min-entropy estimation ($\bar{\epsilon}$) were calculated in the same way as in \cite{HDFK}, with $\epsilon_{\rm EC}=\epsilon_{\rm PA}=\bar{\epsilon}=10^{-10}$ for optimal results \cite{supplemental}.

One complication in the hybrid scheme arises from Alice's switching of the photon sources, which truncates her SPDC's two-photon joint temporal width to the frame duration $T_f$. This causes the first bin to lose Franson visibility because it would have required quantum interference between light from the first bin of the SPDC frame traveling through the short path (of the Franson interferometer) and light from the last bin of the previous PPM frame traveling through the long path.  Figure~\ref{f2}(a) plots the average Franson visibilities measured for different frame sizes. The inset of Fig.~\ref{f2}(a) plots time-bin resolved Franson visibilities for frame size $N=10$, showing near-unity visibilities in all bins except the first one. The residual visibility in the first bin in Fig.~\ref{f2}(a) is due to interference that occurs, at low probability, when the Franson frame is preceded by another Franson frame. In this work we use the maximum time-bin resolved visibility as a measure of the two-photon frequency anti-correlation needed to bound Eve's Holevo information. This treatment circumvents the visibility degradation in the first bin, which does not reflect the intrinsic correlation between the SPDC signal and idler photons. 
For QKD operation, we set an ASE photon rate $\mu_{\text{PPM}}=0.5$, and a low $\mu_{\text{SPDC}}=0.005$ to maintain near-unity experimental Franson visibility of $V/V^{\rm th}=99.7\pm0.1$\% that leads to a small $\chi^{\rm E}$. 

The ASE and SPDC signal light are indistinguishable to Eve, except for the disparity in their mean photon numbers. To defeat Eve's photon-number-splitting attack~\cite{PNS}, we implement a passive decoy-state method~\cite{passivedecoy} by grouping the time bins within an SPDC frame into subgroups to create an adjustable intensity per subgroup~\cite{passdecoy}. In addition to the ASE signal state, we choose two SPDC decoy states to estimate $\underline{Q_{1}}$ and $\overline{\chi^{\rm E}_{1}}$, in which we consider $\epsilon_{\text{decoy}}=10^{-6}$ for the finite data analysis. Unlike conventional decoy-state operation with active modulation~\cite{decoy}, our decoy-state method generates signal/decoy intensity levels passively in the post-processing stage. This removes the requirement of fast QRNGs and significantly reduces the possibility of signal/decoy information leakage in the source~\cite{sourceattacks}. The details of our decoy-state estimation are given in Supplemental Material \cite{supplemental}.

Figure~\ref{f2}(b) plots the optimized PIE results for different frame sizes showing that they increase with increasing $N$. The maximum $\Delta I_{\rm AB}^N$ (after considering the finite key effect) is 4.0 bits per detected photon at $N=128$ with $\chi^{\rm E}=2.1$ bits. The trend of increasing SI at larger $N$ is expected given the precise timing of the generated PPM pattern, for which the dominant causes of symbol errors were afterpulsing of the InGaAs SPADs \cite{Restelli}, i.e., an effect that does not scale up with increasing frame size. Figure~\ref{f2}(c) plots the final secret-key throughput (after decoy-state processing) together with the photon reception rates for different frame sizes. The key throughput exceeds photon-reception rate for all frames larger than $N=4$, a metric no other QKD protocol but PPM-QKD has attained. The peak throughput of 8.0 Mbit/s, higher than the photon detection rate of 6 MHz, is achieved at $N=8$ with a PIE $\Delta I^{8}_{\rm AB}=2.3$ bits per photon ($\chi^{\rm E}=0.37$ bits). This key rate could have been higher but was limited by the TDC maximum recording speed and Bob's sub-optimal basis selection ratio. With optimized ratio and no counter constraint, we could expect a key rate of about 20 Mbit/s. We also point out that the demonstrated PPM-QKD throughput is on par with that of the entanglement-based HDQKD \cite{Zhong}, but with much less reliance on highly efficient single-photon detectors.



\begin{figure}[htb]
\includegraphics[width=0.4\textwidth]{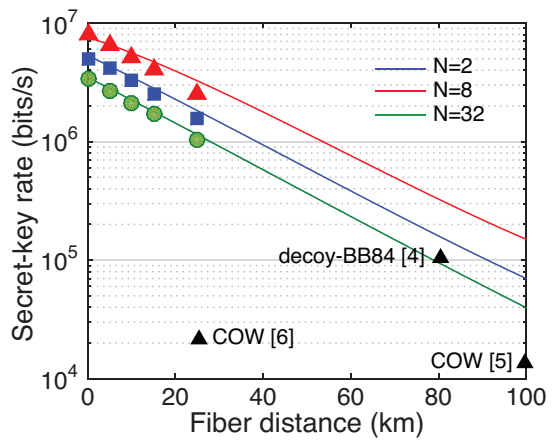}
\caption{(color online). Experimental PPM-QKD throughputs and theoretical predictions. Measured key rates up to loss equivalent to 25 km of fiber propagation are plotted for $N=2$ (squares), $N=8$ (triangles) and $N=32$ (circles). Solid lines are theoretical rates for the same experimental parameters. Data for other frame sizes show similar agreement with the theory, and are not plotted here for better clarity. Reported key rates of other protocols within the range are also plotted.}\label{f3}
\end{figure}

Next, we investigated QKD rates at increasing channel loss up to 5 dB using variable attenuation to simulate 0.2\,dB/km loss due to fiber propagation. Figure~\ref{f3} compares the results for $N=2, 8, 32$ to theoretical calculations that extend to 100 km. The theoretical curves use the same parameters as those used in the experiment and assume the Franson visibilities degrade over distance only due to diminished coincidence-to-accidental ratios. As expected, the measured key rates decrease linearly with distance up to 25\,km, a suitable range for a metropolitan quantum link.  At 25 km, PPM-QKD produces a maximum key rate of 2.5 Mbit/s ($N=8$), and its PIE at 2.3 bits per detection is the same as in short distances. That PIE does not degrade over distances indicates that Eve's Holevo information is virtually independent of Bob's channel loss in the tested range because Franson visibility is highly tolerant to loss and dispersion \cite{ndc}. We did not measure key rates beyond 25 km due to long integration times for Franson measurements; however, high-fidelity Franson statistics can be established at longer distances within a reasonable duration by using high-efficiency detectors \cite{Cuevas}. Note that the projected secret-key rates with $N=8$ remain higher than the photon reception rates up to 100 km, with a value above 100 kbit/s at 100 km.

To further unleash the power of PPM-QKD, we consider plausible future extensions of the protocol. The Franson interferometer delay $\Delta T$ can be made reconfigurable and set equal to the frame duration $T_f$ that varies with $N$. Referring to our security analysis \cite{supplemental}, for a fixed degradation of Franson visibility, a larger $\Delta T$ leads to a lower excess noise of the frequency correlation Eve could exploit. This modification thus allows a tighter bound on $\chi^{\rm E}$ for large frames, therefore boosting the PIE key capacity even beyond the present experiment. With the same experimental parameters used in Fig. 2, except that $\Delta T=T_f$, we obtain a PIE of 8.8 bits for $N=1024$, and a resultant key throughput about 5 times higher than the photon reception rate. In applications in which the transmitter's photon flux is constrained, such as satellite-based sources with a limited power consumption budget, the gain in PIE offered by PPM-QKD directly translates to an enhancement of the key throughput at the receiver.  

To conclude, we have proposed and implemented PPM-QKD that is secure against collective Gaussian attacks via PPM encoding of a classical light source and highly-accurate Franson interferometry using an auxiliary time-energy entangled source. Our proof-of-principle system demonstrates photon-efficient key distribution using practical InGaAs SPADs, achieving 2.5 Mbit/s throughput at loss equivalent to 25 km of fiber that is unambiguously higher than the photon reception rate. We note that our unoptimized PPM-QKD demonstration already delivered secret-key rates comparable to the highest of decoy-state BB84~\cite{shields} with much fewer transmitted photons, and we expect the trend to persist at longer distances because of the same scaling of key rates with channel loss. Unlike decoy-state BB84, PPM-QKD implements the decoy state protocol passively, hence it may be less prone to information leakage. Our PPM scheme may thrive in quantum communication applications that operate in the photon-starved reception regime, such as satellite-based quantum crypto-links with constrained photon-flux transmitters~\cite{entQKD2, freespace} or a quantum local (metropolitan) access network in which end user-nodes use inexpensive detectors with limited detection efficiencies and count rates.

This work was supported in part by the DARPA InPho program under Army Research Office grant number W911NF-10-1-0416 and by the Office of Naval Research grant number N00014-13-0774.

\clearpage

\newcommand{\beginsupplement}{%
        \setcounter{table}{0}
        \renewcommand{\thetable}{S\arabic{table}}%
        \setcounter{figure}{0}
        \renewcommand{\thefigure}{S\arabic{figure}}%
     }

\onecolumngrid

      \beginsupplement
      
      \section{Supplemental Material for photon-efficient quantum cryptography with pulse-position modulation}

\section{Estimating Eve's Holevo information from Franson visibility} \label{Sec:Holevo}
With no access to Alice's idler beam, Eve's interaction can only disturb the frequency variance $\langle\hat{\omega}_{\rm B}^2\rangle$ of the signal light that Bob receives and the frequency covariance $\langle\hat{\omega}_{\rm A}\hat{\omega}_{\rm B}\rangle$ between the conjugate signal and idler measurements. The total change in the mean-squared frequency difference $\langle(\hat{\omega}_{\rm A}-\hat{\omega}_{\rm B})^2\rangle$ due to Eve's intrusion is thus bounded by
\begin{equation}
 \Delta\langle\hat{\omega}_{\rm B}^2-2\hat{\omega}_{\rm A}\hat{\omega}_{\rm B}\rangle\leq2(V^{\rm th}-V)/\Delta T^2,
 \label{eq1}
\end{equation}
where $V^{\rm th}$ is the theoretical Franson visibility for an unperturbed entangled pair assuming a perfect measurement apparatus. Following the proofs for Gaussian CV-QKD protocols based on the optimality of Eve's collective Gaussian attacks for a given time-frequency covariance matrix (TFCM) $\Gamma$ \cite{patron}, Eq.~\eqref{eq1} constrains the set, $\mathcal{M}$, of physically allowed TFCMs with corresponding frequency variance and covariance elements. An upper bound on Eve's Holevo information for infinite key length is then calculated by maximizing $\chi^{\rm \Gamma}=S(\hat{\rho}_{\rm E})-S(\hat{\rho}_{{\rm E}|t_{\rm A}})=S(\hat{\rho}_{\rm AB})-S(\hat{\rho}_{{\rm B}|t_{\rm A}})$ over all TFCMs in $\mathcal{M}$, i.e.,
\begin{equation}
\chi^{\rm E}_{\text{inf}}=\sup_{\Gamma\in \mathcal{M}} \{\chi^{\rm \Gamma}\},
\label{eq2}
\end{equation}
where $\hat{\rho}_{{\rm E}|t_{\rm A}}$ denotes the Eve's quantum state conditioned on Alice's arrival-time measurement, and we assume Alice, Bob and Eve share a pure joint-Gaussian state.

The following gives detailed steps to calculate $\chi^{\rm E}_{\text{inf}}$, which can also be found in \cite{Zhang}. The undisturbed state of one signal-idler photon pair generated from cw SPDC takes the form
\begin{equation}
\label{time-energy}
|\phi\rangle=\iint d t_{\rm A} d t_{\rm B} e^{-\frac{(t_{\rm A}+t_{\rm B})^2}{16\sigma_{\rm coh}^2}} e^{-\frac{(t_{\rm A}-t_{\rm B})^2}{4\sigma_{\rm cor}^2}} e^{-i\omega_{\rm p}\frac{(t_{\rm A}+t_{\rm B})}{2}}|t_{\rm A}\rangle_{\rm A}|t_{\rm B}\rangle_{\rm B},
\end{equation}
where $\sigma_{\rm coh}$ is the pump coherence time, and $\sigma_{\rm cor}$ is the biphoton correlation time. The two photons are correlated in the time domain, and anti-correlated in the frequency domain where time and frequency form a pair of conjugate bases. We thus introduce the arrival-time operator $\hat{t}_m$ and the frequency operator $\hat{\omega}_n$, where $m, n \in\{A, B\}$. The state $|\omega\rangle_{\rm A} (|\omega\rangle_{\rm B})$ in this paper represents a single photon of the signal (idler) at frequency $\omega_{\rm p}/2+\omega$ $(\omega_{\rm p}/2-\omega)$, so that with this convention the detunings, $\omega$, from $\omega_p/2$ are correlated, rather than anti-correlated. The TFCM for the above state is then
\begin{equation}
\Gamma^0= \begin{bmatrix} \gamma^0_{\rm AA} & \gamma^0_{\rm AB} \\\gamma^0_{\rm BA}  & \gamma^0_{\rm BB} \end{bmatrix},
\end{equation}
where
\begin{eqnarray}
\label{AB0}
\gamma^0_{\rm AA}=\gamma^0_{\rm BB} &=& \begin{bmatrix} \frac{1}{4}\sigma_{\rm cor}^2+\sigma_{\rm coh}^2 & 0 \\0  & \frac{1}{4\sigma_{\rm cor}^2}+\frac{1}{16\sigma_{\rm coh}^2} \end{bmatrix} \nonumber\\
\gamma^0_{\rm AB}=\gamma^0_{\rm BA} &=& \begin{bmatrix} -\frac{1}{4}\sigma_{\rm cor}^2+\sigma_{\rm coh}^2 & 0 \\0  & \frac{1}{4\sigma_{\rm cor}^2}-\frac{1}{16\sigma_{\rm coh}^2} \end{bmatrix}.
\end{eqnarray}

Eve's presence disturbs Alice and Bob's initial TFCM to become
\begin{eqnarray}
\label{AB}
\gamma_{\rm AA} &=&\gamma^0_{\rm AA} \nonumber\\
\gamma_{\rm AB}=\gamma_{\rm BA}&=& \begin{bmatrix} 1-\eta_t & 0 \\0  & 1-\eta_{\omega} \end{bmatrix}\gamma^0_{\rm AB} \nonumber\\
\gamma_{\rm BB}&=& \begin{bmatrix} 1+\epsilon_t & 0 \\0  & 1+\epsilon_{\omega}\end{bmatrix}\gamma^0_{\rm BB},
\end{eqnarray}
where $\{\eta_t,\eta_{\omega}\}$ denotes the loss in time and frequency correlation, and $\{\epsilon_{t},\epsilon_{\omega}\}$ denotes the excess noise in Bob's photon. The measured Franson visibility restricts the possible $\eta_{\omega}$, $\epsilon_{\omega}$ values via inequality (1) in the main text. We note that any disturbance in the biphoton time correlation or Bob's arrival time variance (reflected by $\eta_t$ and $\epsilon_t$) cannot be bounded by our Franson interference measurement. Nevertheless, such disturbance by Eve does not afford her any benefit in gaining symbol information encoded in the time basis, thus it has negligible impact on $\chi^{\rm E}_{\text{inf}}$. To ensure stronger security, we therefore take the mean-squared time of arrival difference $\langle(\hat{t}_{\rm A}-\hat{t}_{\rm B})^2\rangle$ to be the square of the detector timing jitter (beyond which Eve's intrusion would have been readily detected by Alice and Bob), and $\langle\hat{t}_{\rm B}^2\rangle$ to be the time variance integrated over the entire frame duration.

For a given TFCM, a Gaussian attack maximizes Eve's Holevo information by assuming that she purifies the state to a joint Gaussian state between Alice, Bob, and Eve.  The Holevo information $\chi^{\rm \Gamma}$ for covariance matrix $\Gamma$ is
\begin{equation}
\label{eve}
\chi^{\rm \Gamma}=S(\hat{\rho}_{\rm E})-\int dt\, p(t_{\rm A}) S(\hat{\rho}_{{\rm E}|t_{\rm A}}),
\end{equation}
where $S(\hat{\rho})=-\textrm{Tr}[\hat{\rho}\log_2(\hat{\rho})]$ is the von Neumann entropy of the quantum state $\hat{\rho}$. Under the assumption that Alice, Bob and Eve's joint quantum state is pure, we have $S(\hat{\rho}_{\rm E})=S(\hat{\rho}_{\rm AB})$ and $S(\hat{\rho}_{{\rm E}|t_{\rm A}})=S(\hat{\rho}_{{\rm B}|t_{\rm A}})$. Furthermore, because all states are Gaussian, the von Neumann entropy of Bob and Eve's conditional quantum state is independent of Alice's measurement result. Thus the integral in Eq.~(\ref{eve}) can be reduced to
\begin{equation}
\label{eve2}
\chi^{\rm \Gamma}=S(\hat{\rho}_{\rm AB})-S(\hat{\rho}_{{\rm B}|t_{\rm A}}).
\end{equation}
To evaluate $\chi^{\rm E}_{\text{inf}}$ from Eq.~\eqref{eve2}, we follow the standard formalism by defining
\begin{eqnarray}
\label{Srho}
I_1 &=& \langle \Delta \hat{t}^2_{\rm A}\rangle \langle \Delta \hat{t}^2_{\rm B}\rangle \nonumber\\
I_2 &=& \langle\Delta \hat{\omega}^2_{\rm A}\rangle \langle\Delta \hat{\omega}^2_{\rm B}\rangle\nonumber\\
I_3 &=&\langle\Delta \hat{t}_{\rm A}\Delta \hat{t}_{\rm B}\rangle \langle\Delta \hat{\omega}_{\rm A}\Delta \hat{\omega}_{\rm B}\rangle\nonumber\\
I_4 &=&(\langle\Delta \hat{t}^2_{\rm A}\rangle \langle\Delta \hat{t}^2_{\rm B}\rangle-\langle\Delta \hat{t}_{\rm A}\Delta \hat{t}_{\rm B}\rangle^2)(\langle\Delta \hat{\omega}^2_{\rm A}\rangle \langle\Delta \hat{\omega}^2_{\rm B}\rangle-\langle\Delta \hat{\omega}_{\rm A}\Delta \hat{\omega}_{\rm B}\rangle^2)\nonumber\\
d_\pm &=&\frac{1}{\sqrt{2}}\sqrt{I_1+I_2+2I_3\pm \sqrt{(I_1+I_2+2I_3)^2-4I_4}}.
\end{eqnarray}
 Then $S(\hat{\rho}_{\rm AB})=f(d_+)+f(d_-)$, where
 \begin{equation}
 f(d)=(d+1/2)\log_2(d+1/2)-(d-1/2)\log_2(d-1/2),
 \end{equation}
 and we have
 \begin{equation}
 S(\hat{\rho}_{{\rm B}|t_{\rm A}})=f(\sqrt{\textrm{det}[\gamma_{{\rm I}|t_{\rm A}}]}),
 \end{equation}
 where Bob's conditional covariance matrix is

\begin{equation}
\label{Bcon}
\gamma_{{\rm I}|t_{\rm A}}= \begin{bmatrix} \langle\Delta \hat{t}^2_{\rm B}\rangle-\langle\Delta \hat{t}_{\rm A}\Delta \hat{t}_{\rm B}\rangle^2/\langle\Delta \hat{t}^2_{\rm A}\rangle & 0 \\0 &\langle\Delta\hat{\omega}^2_{\rm B}\rangle \end{bmatrix}.
\end{equation}
Our upper bound on Eve's Holevo information for infinite key length is then found from
\begin{equation}
\chi^{\rm E}_{\text{inf}}=\sup_{\Gamma\in \mathcal{M}} \{\chi_{\rm \Gamma}\}\nonumber.
\end{equation}

\section{Calculation of information loss due to finite key lengths}
\noindent The finite-data analysis against collective attacks follows~\cite{HDFK}. For the estimation of $\chi^{\rm E}$, Eve's Holevo information with finite key consideration, Alice and Bob calculate their normalized frequency correlation from measured Franson visibilities via Eq.~(\ref{eq1}), which has a $\chi^2$ distribution:
\begin{equation}
(m-1)\frac{\langle(\hat{\omega}_{\rm A}-\hat{\omega}_{\rm B})^2\rangle}{\langle(\hat{\omega}_{\rm A0}-\hat{\omega}_{\rm B0})^2\rangle} \sim \chi^2(1-\epsilon_{PE},m-1),
\end{equation}
where $m$ is the number of Franson visibility measurements taken. An upper bound on $\langle(\hat{\omega}_{\rm A}-\hat{\omega}_{\rm B})^2\rangle$ with confidence interval $1-\epsilon_{PE}$ is then given by:
\begin{equation}
\langle(\hat{\omega}_{\rm A}-\hat{\omega}_{\rm B})^2\rangle_{\rm max}=
\langle(\hat{\omega}_{\rm A0}-\hat{\omega}_{\rm B0})^2\rangle + \frac{2}{\sqrt{m}}{\rm erf}^{-1}(1-\epsilon_{PE})\langle(\hat{\omega}_{\rm A}-\hat{\omega}_{\rm B})^2\rangle.
\end{equation}
This upper bound is then used to calculate the worst case $\chi^{\rm E}$ and the most pessimistic secure PIE based on the procedure in section I. In our experiment, $m=100$, and  we choose $\epsilon_{PE}=10^{-5}$. The overall failure probability of the entire protocol is thus $\epsilon_s=\epsilon_{EC}+\epsilon_{PA}+\bar{\epsilon}+\epsilon_{PE}\approx10^{-5}$.

\section{Theoretical model} \label{Sub:Model:ChDet}
In this section, we present the model used to obtain the theoretical rates presented in Fig. 3 of the main text. We use the proposal from~\cite{Norbert} to model the system. We use a beam splitter with transmissivity $\eta$ followed by an ideal single-photon detector with no photon number resolution capability to model the channel and detection. The system transmissivity $\eta$ is given by:
\begin{equation}\label{Model:Eta}
\eta=\eta_{B}10^{-\alpha l/10},
\end{equation}
where $\eta_{B}$ denotes the transmissivity on Bob's side, including the internal transmission efficiency of optical components and detector efficiency. The channel loss is included for a transmission distance $l$ km with a loss coefficient $\alpha$ measured in dB/km. Then, the efficiency $\eta_i$ of the $i$-photon state with respect to a detector is given by:
\begin{equation}\label{Model:etai}
\eta_i=1-(1-\eta)^i
\end{equation}
for $i=0,1,2,\cdots$.

\textbf{Yield:} Define $Y_i$ as the yield of an $i$-photon state, i.e., the conditional probability of a detection event by Bob, given that Alice sends an $i$-photon state. Note that $Y_0$ is the background rate which includes detector dark counts and other background contributions. The yield of the $i$-photon states $Y_i$ comes mainly from two parts, the background and the true signal. Assuming that the
background counts are independent of the signal photon detection, then $Y_i$ is given by:
\begin{equation}\label{Model:Yi}
\begin{aligned}
Y_i &= Y_0 + \eta_i - Y_0\eta_i \cong Y_0 + \eta_i.
\end{aligned}
\end{equation}
Here, we assume $Y_0$ (typically $10^{-5}$) and $\eta$ (typically $10^{-3}$) are small.

\textbf{Gain:} The gain of $i$-photon states $Q_i$ is given by:
\begin{equation}\label{Model:Qi}
\begin{aligned}
Q_i &= Y_i\frac{\mu^i}{i!}e^{-\mu},
\end{aligned}
\end{equation}
where $\mu$ is defined as the mean photon number \emph{per frame}. The gain $Q_i$ is the probability that Alice sends out an $i$-photon state and Bob obtains a detection in a frame. Then the overall gain, the probability for Bob to obtain a detection event in a frame, is the sum over all $Q_i$'s:
\begin{equation}\label{Model:Gain}
\begin{aligned}
Q_{\mu} &= \sum_{i=0}^{\infty} Y_i\frac{\mu^i}{i!}e^{-\mu}. \\
\end{aligned}
\end{equation}

In the QKD scenario that we are considering, Eve's attack can change the $Y_i$. Without Eve, in normal QKD, Eqs.~\eqref{Model:etai}--\eqref{Model:Qi} are satisfied for all $i=0,1,2,\cdots$. Thus, the gain is given by:
\begin{equation}\label{Model:WithoutEve}
\begin{aligned}
Q_{\mu} &= Y_0 + 1-e^{-\eta\mu}. \\
\end{aligned}
\end{equation}
Due to the fact that $Q_\mu$ can be measured or tested experimentally, we will use Eq.~\eqref{Model:WithoutEve} in our theoretical model.

Now, let Alice's optical switch have a probability $P_{\text{os}}$ in choosing the ASE source. If the frame length is $T_{f}$, then the expected count rate for the ASE source in the timing measurement is given by
$$
N_{c}=\frac{1}{T_{f}}P_{\text{os}}Q_{\mu},
$$
where we assume that there is no limitation on the count rates of the TDC.

\section{Decoy-state analysis} \label{Sec:decoy}
We show how to use the decoy-state method to obtain the lower bound on the single-photon gain $\underline{Q_1}$ and the upper bound on the single-photon mean-squared frequency difference $\overline{\langle(\hat{\omega}_{\rm A}-\hat{\omega}_{\rm B})^2\rangle_{1}}$. To generate different intensity levels of decoy states, we group the time bins within an SPDC frame into subgroups~\cite{passdecoy}. That is, Bob partitions an SPDC frame into two subgroups containing $F_{\nu_1}$ ($F_{\nu_2}$) bins in the first (second) subgroup, with a corresponding mean photon number per subgroup $\nu_1$ ($\nu_2$). For instance, for an SPDC frame with $N$ time bins, we define the initial mean photon number per frame as $\nu=N\nu_{\text{bin}}$, where $\nu_{\text{bin}}$ is the mean photon number per bin. We can choose $F_{\nu_1}=N/2$ and $F_{\nu_2}=N/4$, then we have two decoy states with $\nu_1=\nu/2$ and $\nu_2=\nu/4$. The signal state $\mu$ is the mean photon number per frame, generated from the ASE source. Once the gains of signal/decoy states are experimentally determined, we can follow~\cite{Darius} for decoy-state estimation where we assume that Alice and Bob choose $\nu_1$ and $\nu_2$ satisfying
\begin{equation}\label{Decoy:ConditionWeak}
\begin{aligned}
0\le\nu_2<\nu_1\\
\nu_1+\nu_2<\mu.
\end{aligned}
\end{equation}

\subsection{Single-photon yield}
The gains of the two decoy states are given by
\begin{eqnarray}
Q_{\nu_1} &=& \sum_{i=0}^{\infty}Y_i\frac{\nu_1^i}{i!} e^{-\nu_1},
\label{Decoy:WeGain1}\\
Q_{\nu_2} &=& \sum_{i=0}^{\infty}Y_i\frac{\nu_2^i}{i!} e^{-\nu_2}.
\label{Decoy:WeGain2}
\end{eqnarray}

First Alice and Bob can estimate the lower bound on the background rate $Y_0$ from
$$
\begin{aligned}
\nu_1Q_{\nu_2}e^{\nu_2}-\nu_2Q_{\nu_1}e^{\nu_1} &=(\nu_1-\nu_2)Y_0-\nu_1\nu_2\sum_{i=1}^{\infty}Y_{i+1}\frac{\nu_1^i-\nu_2^i}{(i+1)!} \le (\nu_1-\nu_2)Y_0.
\end{aligned}
$$
Thus, a lower bound of $Y_0$ is given by
\begin{equation}\label{Decoy:Y0Low}
\begin{aligned}
Y_0 \ge \underline{Y_0} = \max\{
\frac{\nu_1Q_{\nu_2}e^{\nu_2}-\nu_2Q_{\nu_1}e^{\nu_1}}{\nu_1-\nu_2},0\}.
\end{aligned}
\end{equation}

Now, from Eq.~\eqref{Model:Gain}, the contribution from multi-photon states (with photon number $\ge2$) in the signal state can be expressed as,
\begin{equation}\label{Decoy:MulGain}
\begin{aligned}
\sum_{i=2}^{\infty}Y_i\frac{\mu^i}{i!} &= Q_\mu e^\mu-Y_0-Y_1\mu.\\
\end{aligned}
\end{equation}

Combining Eqs.~\eqref{Decoy:WeGain1} and \eqref{Decoy:WeGain2}, under condition Eq.~\eqref{Decoy:ConditionWeak}, we have~\cite{MaDecoy}
\begin{equation}\label{Decoy:WeGainBound}
\begin{aligned}
&Q_{\nu_1}e^{\nu_1}-Q_{\nu_2}e^{\nu_2}= Y_1(\nu_1-\nu_2) + \sum_{i=2}^{\infty}\frac{Y_i}{i!}(\nu_1^i-\nu_2^i) \le Y_1(\nu_1-\nu_2) + \frac{\nu_1^{2}-\nu_2^2}{\mu^{2}}\sum_{i=2}^{\infty}Y_i\frac{\mu^{i}}{i!} \\
& = Y_1(\nu_1-\nu_2) + \frac{\nu_1^{2}-\nu_2^2}{\mu^{2}}(Q_\mu e^\mu-Y_0-Y_1\mu) \le Y_1(\nu_1-\nu_2) + \frac{\nu_1^{2}-\nu_2^2}{\mu^{2}}(Q_\mu e^\mu-\underline{Y_0}-Y_1\mu),\\
\end{aligned}
\end{equation}
where $\underline{Y_0}$ is defined in Eq.~\eqref{Decoy:Y0Low}. By solving inequality \eqref{Decoy:WeGainBound}, the lower bound of
$Y_1$ is given by
\begin{equation}\label{Decoy:Y1Bound}
\begin{aligned}
& Y_1 \ge \underline{Y_1} =
\frac{\mu}{\mu\nu_1-\mu\nu_2-\nu_1^2+\nu_2^2}\left[Q_{\nu_1}e^{\nu_1}-Q_{\nu_2}e^{\nu_2}-\frac{\nu_1^2-\nu_2^2}{\mu^2}(Q_\mu
e^\mu-\underline{Y_0})\right].
\end{aligned}
\end{equation}

\subsection{Single-photon frequency correlation}
As mentioned in Sec.~\ref{Sec:Holevo}, the mean-squared frequency difference $\langle(\hat{\omega}_{\rm A}-\hat{\omega}_{\rm B})^2\rangle$ can be bounded by the Franson visibility. We have that $\langle(\hat{\omega}_{\rm A}-\hat{\omega}_{\rm B})^2\rangle$ can be written as
\begin{eqnarray} \label{Eq:Deltanunu}
\langle(\hat{\omega}_{\rm A}-\hat{\omega}_{\rm B})^2\rangle &=& \frac{Q_1}{Q_{\nu}}\langle(\hat{\omega}_{\rm A}-\hat{\omega}_{\rm B})^2\rangle_{1}+(1-\frac{Q_1}{Q_{\nu}})\langle(\hat{\omega}_{\rm A}-\hat{\omega}_{\rm B})^2\rangle_{m},
\end{eqnarray}
where $\langle(\hat{\omega}_{\rm A}-\hat{\omega}_{\rm B})^2\rangle_{1}$ is the mean-squared frequency difference due to single-pair emission, and $\langle(\hat{\omega}_{\rm A}-\hat{\omega}_{\rm B})^2\rangle_{m}$ is the mean-squared frequency difference due to multiple pair emissions and dark counts. We can then upper bound $\langle(\hat{\omega}_{\rm A}-\hat{\omega}_{\rm B})^2\rangle_{1}$ as follows:
\begin{equation} \label{Decoy:Delta1Bound}
\begin{aligned}
& \langle(\hat{\omega}_{\rm A}-\hat{\omega}_{\rm B})^2\rangle_1 \leq \overline{\langle(\hat{\omega}_{\rm A}-\hat{\omega}_{\rm B})^2\rangle_1} =
\frac{Q_{\nu}\langle(\hat{\omega}_{\rm A}-\hat{\omega}_{\rm B})^2\rangle}{\underline{Y_1}\nu e^{-\nu}},
\end{aligned}
\end{equation}
where $\nu=N\nu_{\text{bin}}$ is the mean photon pairs per SPDC frame.

\subsection{Finite-data effect}
In the context of Gaussian attacks in our security analysis, we adopt the standard error analysis~\cite{MaDecoy} to estimate the finite-key effect in decoy-state operation. The lower bound and upper bound on the gains are given by
\begin{eqnarray} \label{statistical fluctuation}
& \overline{Q_{\lambda}}=Q_{\lambda}(1+\frac{n_{\alpha}}{\sqrt{N_{\lambda}Q_{\lambda}}}),\\
& \underline{Q_{\lambda}}=Q_{\lambda}(1-\frac{n_{\alpha}}{\sqrt{N_{\lambda}Q_{\lambda}}}),
\end{eqnarray}
where $\lambda\in\{\mu,\nu_1,\nu_2\}$, $n_{\alpha}$ is the number of standard deviations one chooses for statistical fluctuation analysis, $N_{\lambda}$ is the number of frames sent by Alice. We can insert these bounds into Eqs.~\eqref{Decoy:Y0Low} and ~\eqref{Decoy:Delta1Bound} to obtain $\underline{Y_1}$ and $\overline{\langle(\hat{\omega}_{\rm A}-\hat{\omega}_{\rm B})^2\rangle_1}$ in the finite-data case.

Finally, the gain of single-photon state is, according to Eq.~\eqref{Model:Qi},
\begin{equation}\label{Decoy:Q1Bound}
\begin{aligned}
Q_1 \ge \underline{Q_1} = \mu e^{-\mu} \underline{Y_1}.
\end{aligned}
\end{equation}
The upper bound on Eve's Holevo information, $\overline{\chi^{\rm E}_{1}}$, can be obtained by using Eq.~\eqref{eq2}, in which the set $\mathcal{M}$ is constrained by $\overline{\langle(\hat{\omega}_{\rm A}-\hat{\omega}_{\rm B})^2\rangle_1}$~\cite{Darius}.

\end{document}